\documentclass[aps,amsfonts,pre,twocolumn,superscriptaddress,showpacs,eqsecnum]{revtex4-1}

\usepackage[normalem]{ulem}
\usepackage{float}
\usepackage[usenames]{color}
\usepackage{graphicx}
\usepackage{ulem}
\usepackage{amsfonts}
\usepackage{amsmath}
\usepackage{natbib,,subfigure}
\usepackage[caption=false]{subfig} 
\usepackage{epsfig,amsmath,amssymb,bm,epsf,psfrag,verbatim}
\usepackage{units} 

\def\Prob{p}

\begin{document}

\title{Nonlinear elasticity of disordered fiber networks}

\author{Jingchen Feng}
\affiliation{Bioengineering Department and Center for Theoretical Biological Physics, Rice University, Houston TX, 77251-1892,USA. }

\author{Herbert Levine}
\affiliation{Bioengineering Department and Center for Theoretical Biological Physics, Rice University, Houston TX, 77251-1892,USA. }

\author{Xiaoming Mao}
\affiliation{Department of Physics, University of Michigan, Ann Arbor, MI 48109-1040, USA. }

\author{Leonard M. Sander}
\affiliation{Physics \& Complex Systems, University of Michigan, Ann Arbor MI 48109-1040, USA.  }

\date{\today}
\begin{abstract}
Disordered biopolymer gels have striking mechanical properties including strong nonlinearities. In the case of athermal gels (such as collagen-I) the nonlinearity has long been associated with a crossover from a bending dominated to a stretching dominated regime of elasticity. The  physics of this crossover is related to the existence of a central-force isostatic point and to the fact that for most gels the bending modulus is small. This crossover induces scaling behavior for the elastic moduli. In particular, for linear elasticity such a scaling law has been demonstrated [Broedersz \textit{et al. Nature Physics}, 2011 \textbf{7}, 983]. In this work we generalize the scaling to the nonlinear regime with a two-parameter scaling law involving three critical exponents. We test the scaling law numerically for two disordered lattice models, and find a good scaling collapse for the shear modulus in both the linear and nonlinear regimes.  We compute all the  critical exponents for the two lattice models and discuss the applicability of our results to real systems.
\end{abstract}

\maketitle


\section{Introduction}
Collagen is the single most abundant protein in the animal kingdom; most of it occurs in the form of fibrous collagen such as collagen-I. A major component of tissue is  disordered networks of fibrous collagen, i.e. collagen polymer gels. These networks are one of many important biological examples of  ``athermal" biopolymer gels -- athermal means that the coiling of the stiff fibrils due to thermal fluctuations is negligible. 
These materials have remarkable mechanical properties: even though the constituents are  in the linear elastic regime (up to the yield point of the network), the network itself is nonlinear  and strain-stiffening. For example, in dilute collagen-I gels the shear modulus increases by orders of magnitude as the strain increases \cite{Roeder,Stein,Storm2005,Broedersz,Broedersz2014}. 
Such nonlinear effects are important in biology, for example in the stiffening and alignment of tissue near a growing tumor \cite{Provenzano}.

The unusual mechanical properties of collagen are believed to be intimately connected to two facts.  First, the fibrils have a large stretching modulus, $k$, and a small bending modulus, $\kappa$. More precisely, we define $\bar \kappa \equiv\kappa/ka^2 $ where $a$ is the mesh size of the network;  typically $\bar{\kappa} \ll1$. For example, for collagen-I  
 we can think of the fibrils as elastic rods. In the dilute case (densities of order 2mg/ml) $\bar \kappa \sim (d/a)^2 \sim 10^{-3}$ where $d \sim 30 nm$ is the diameter of the fibrils and $a \sim 1 \mu$ is the mesh size.
Second, the disordered structure of collagen can accommodate small deformations through bending the fibrils, but at large deformation, the bending modes are exhausted and the fibrils have to be stretched.  Owing to the huge difference between the bending and stretching stiffness of the fibrils, such a bending-to-stretching crossover lead to a dramatic increase of the elastic moduli\cite{Onck,Wyart,Sheinman2012,Feng2015,Alexander1998}. 

The existence of such a bending-to-stretching crossover is controlled by the physics of the central-force isostatic point (CFIP).  The concept of CFIP traces back to J. C. Maxwell~\cite{Maxwell1864}. He pointed out that the onset of mechanical instability for a system of particles with central force interactions occurs when the mean coordination number of bonds is twice the spatial dimension, $\langle z\rangle =2d$. At this point the number of degrees of freedom, $d$, and the number of constraints, $z/2$, of each particle are the same~\cite{Liu2010,Mao2010,Mao,Ellenbroek2011,Mao2015,Rocklin2014,Zhang2015a,Lubensky2015}.  For $\langle z\rangle< 2d$ there are soft floppy modes;  for $\langle z\rangle > 2d$ there are redundant bonds and states of self-stress.  Thus the CFIP is a critical point of a mechanical phase transition.  Biopolymer gels with $\bar \kappa  \ll 1$  can be thought of as a nearly central-force system if we identify  the crosslinking points as the particles and the  polymer fibrils between them as bonds.  The coordination number $z$ depends on the type of gel.  Each crosslinking point can have a maximum coordination number $z_{\textrm{max}}$, depending on the type of bonding.  The most common situation, two polymers meeting at a crosslinking point, corresponds to $z_{\textrm{max}}=4$ (Collagen I belongs to this class).  The average coordination number $\langle z\rangle < z_{\textrm{max}}$ because of dangling ends.  Therefore, pure Collagen I and similar biopolymer gels have mean coordination $\langle z\rangle < 2d$ and are always below the CFIP.  In contrast, if a gel has crosslinking points with $z_{\textrm{max}}>2d$ the gel can go above the CFIP when the filaments are long. This may be the case for  multi-component complex gels seen in extracellular matrix material.

\begin{figure}[!ht]
	\centering
		\subfigure[]{\includegraphics[width=.47\textwidth, height=0.22\textwidth]{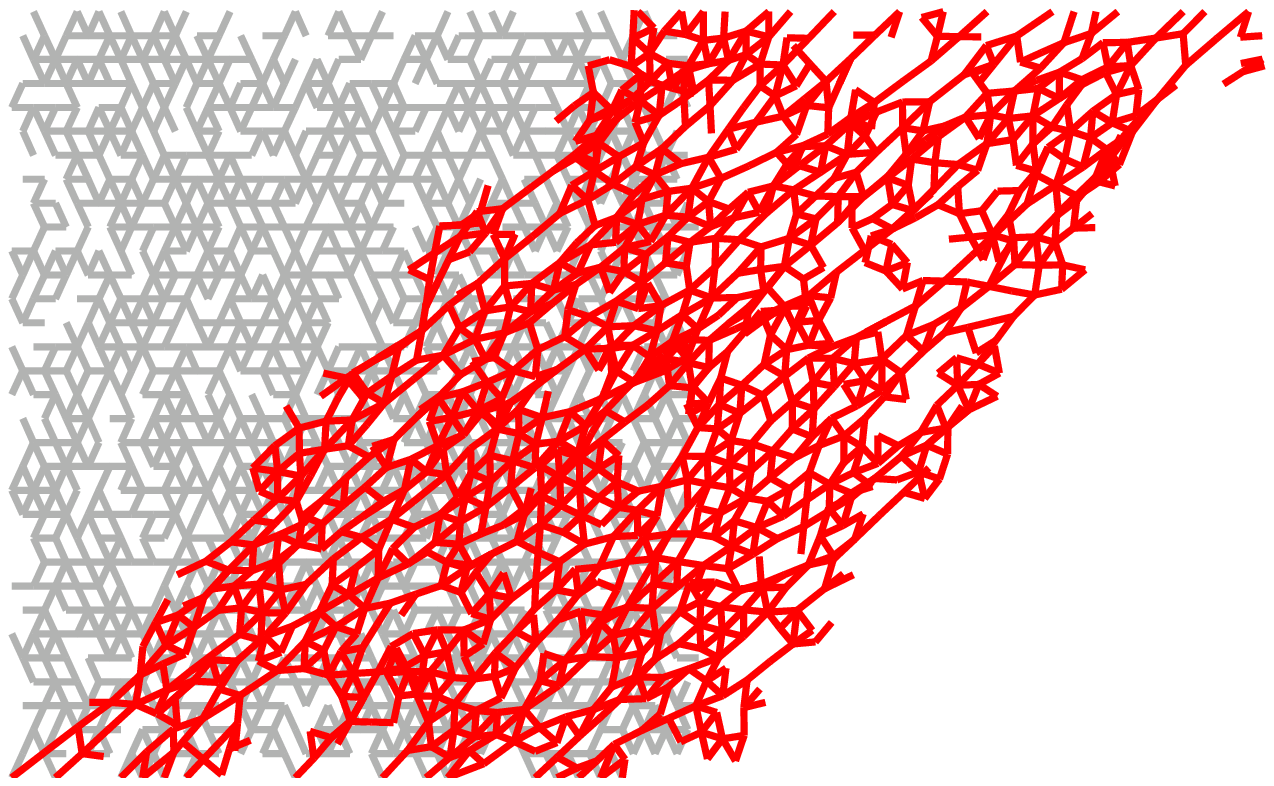}}
		\subfigure[]{\includegraphics[width=.47\textwidth, height=0.22\textwidth]{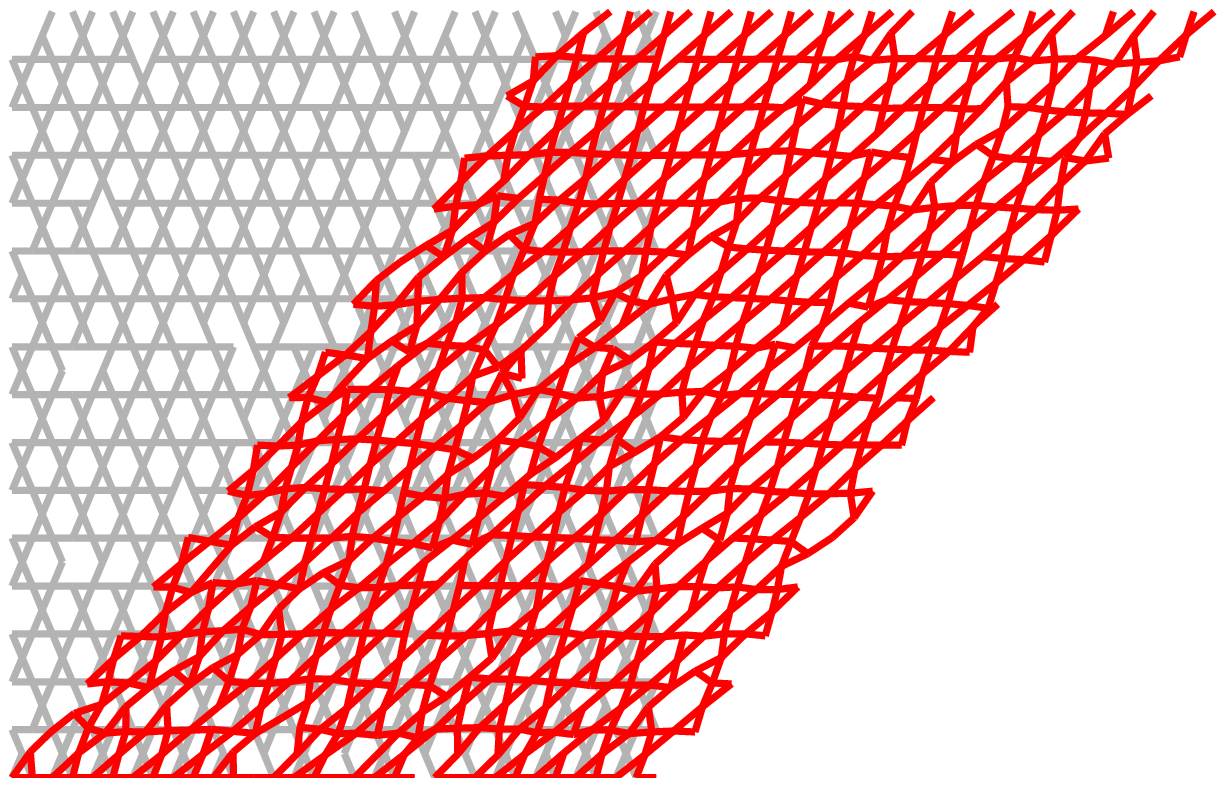}}
		\subfigure[]{\includegraphics[width=.35\textwidth]{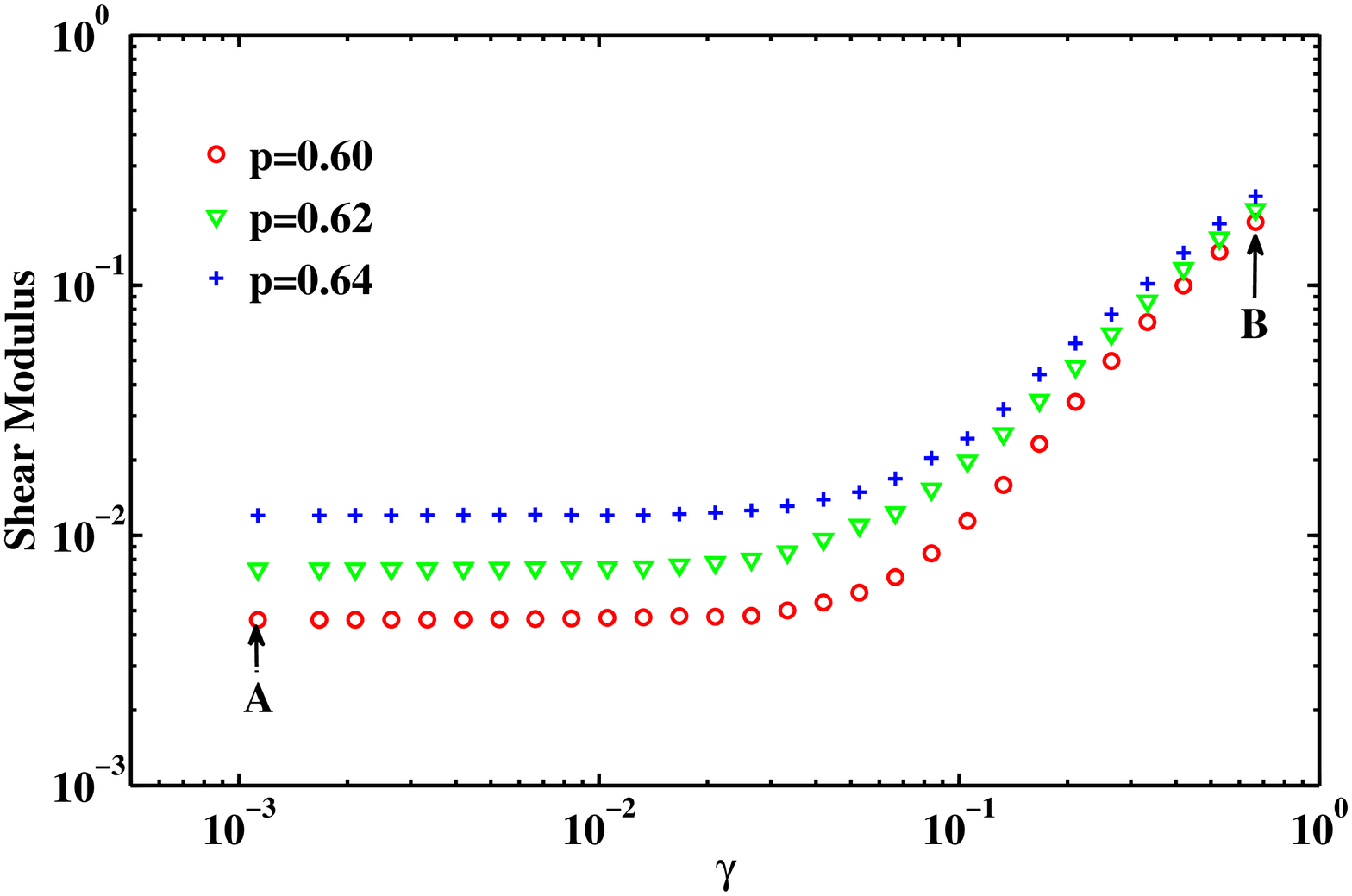}}
		\subfigure[]{\includegraphics[width=.35\textwidth]{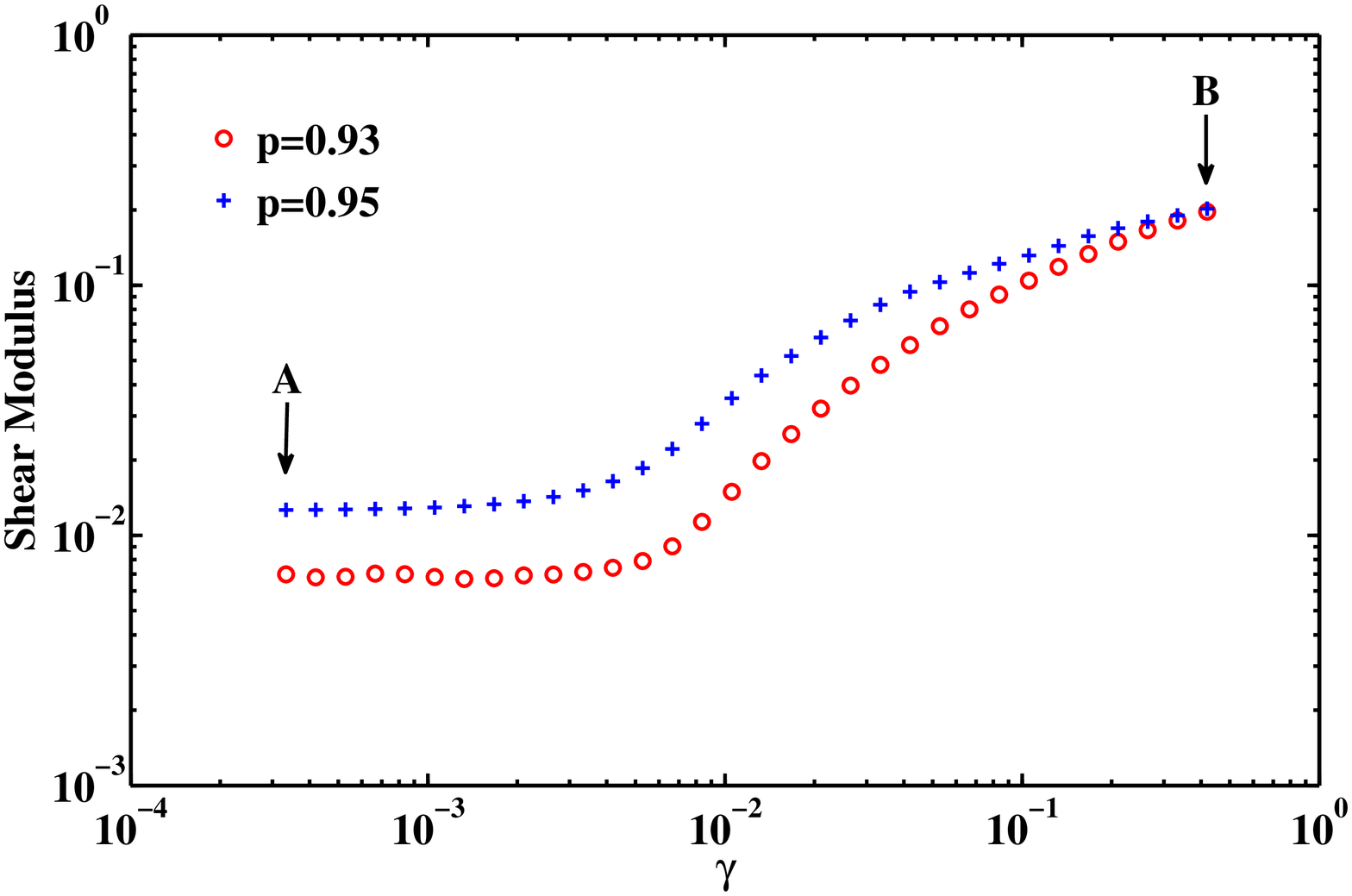}}
		\caption{\label{FIG:lattices} Examples of network configurations before (grey) and after (red) a large deformation for the diluted triangular lattice  at $ \bar{\kappa}=10^{-3}$ (a) and kagome lattice at $ \bar{\kappa}=10^{-4}$. (b). Shear modulus as a function of strain for varying values of $\Prob$ (see legends) for the diluted triangular lattice (c) and kagome lattice (d). Point A(B) corresponds to the gray(red) configuration in Fig 1ab.
		}. 
\end{figure}

Since the  CFIP is a critical point, it is natural to suspect that mechanical properties such as the shear modulus would exhibit scaling properties in its vicinity, and, indeed, for linear elasticity such scaling has been demonstrated \cite{Broedersz,Mao2013b,Mao2013c}. 
In this paper we extend scaling relations near the CFIP  to the  nonlinear regime. As simplified models of real biopolymers we use simulations on two different disordered lattice models, the diluted triangular and kagome lattices (see Fig.~\ref{FIG:lattices} (a,b))\cite{Broedersz,Mao2013b,Mao2013c}.  These two lattices have maximum coordination $z_{\textrm{max}}=6$ and $z_{\textrm{max}}=4$; they represent  two classes of gels with $z_{\textrm{max}}>2d$ and $z_{\textrm{max}}\le 2d$.  We use these lattices to verify our scaling laws and to extract exponents.  We discuss the application of the scaling regimes to more realistic  biopolymer networks in Sec.~\ref{SEC:ConcDisc}.

The scaling form that has been proposed~\cite{Broedersz,Mao2013b,Mao2013c} previously  for the linear shear modulus, $G$ is:
\begin{align}\label{EQ:GLinear}
	G(\Delta z,\bar\kappa) = k\vert \Delta z \vert^{f} \,\mathcal{G}_{\pm ,\textrm{linear}} \left( {\bar \kappa /\vert \Delta z \vert^{\phi}} \right) .
\end{align}
Here $\Delta z \equiv \langle z\rangle -2d$ is the extra coordination above the CFIP and $\mathcal{G}$ is a scaling function.  The exponent $f$ captures the emergence of rigidity as coordination increases in the central-force system, and the crossover exponent $\phi$ captures the effect of $\bar \kappa$ as a relevant perturbation near isostaticity.  This  scaling law for the  linear elasticity includes the  observed effect that the elastic modulus of the gel can change dramatically when the connectivity is changed near the CFIP.  Below the CFIP the elasticity of the gel is bending dominant, $G\sim \kappa$, because it is floppy in the central-force limit.  Above the CFIP the gel becomes stretching dominant $G\sim k$.  If $z_{\textrm{max}}>2d$ there is a special regime where bending and stretching modes are strongly coupled, $G\sim \kappa^{f/\phi} k^{1-f/\phi}$.  This law has been shown to be valid using two and three dimensional disordered lattice models~\cite{Broedersz,Mao2013b,Mao2013c}.

We now turn to the nonlinear elasticity. A generalized scaling law should describe the very large strain-stiffening that is characteristic of these materials. As we discussed above, the physical origin of this effect is that  gels  in biological systems  are often below the CFIP ($\langle z \rangle <2d$).  In  the linear elasticity regime the shear modulus is proportional to the bending stiffness, $\kappa$, and is very small because the shear is accommodated in the floppy bending modes.  As the deformation progress, the soft modes are stretched out and central-force paths form to bear the stress, increasing $G$\cite{Onck}.  

Based on these ideas, we propose a two-parameter scaling law for the differential shear modulus:
\begin{align}\label{EQ:GFull}
	G(\Delta z,\bar \kappa,\gamma) = k\vert \Delta z \vert^{f} \,\mathcal{G}_{\pm} \left( \frac{\bar \kappa }{\vert \Delta z \vert^{\phi}} , \frac{\gamma}{\vert \Delta z \vert^{\beta}}\right) ,
\end{align}
Here $\gamma$ is the shear strain, and the $\pm$ subscript labels the branches of the scaling function for $\Delta z>0$ and $\Delta z<0$, respectively.  This scaling law captures the stiffening effect of both the bending rigidity, $\kappa$, and of the strain, $\gamma$.  

A one-parameter version of this scaling relation for the nonlinear case has been previously proposed by Wyart et al~\cite{Wyart}. These authors considered  varying $\gamma$ with fixed $\kappa$. They  give numerical evidence  that $\beta=1$ for jammed packings, and demonstrated the scaling of  the crossover from linear to nonlinear behavior. 
Our goal is to advance such scaling relations to include both the connectivity $z$ and the dimensionless bending stiffness $\bar \kappa$ as variables. We will collapse the whole elastic modulus curve with varying $z$ and $\bar \kappa$, and analyze different elastic regimes.  Our results are summarized in Figs.~\ref{FIG:Final} and ~\ref{FIG:PD}.

\section{Models and Simulations}
The model lattices are the diluted triangular~\cite{Das2007,Broedersz,Das2012} and kagome lattices~\cite{Mao,Mao2013c}, as shown in Figure \ref{FIG:lattices}ab.  In both lattices each bond is present with probability $\Prob$. The undiluted triangular and kagome lattices have coordination numbers $z_{\textrm{max}}=6$ and $z_{\textrm{max}}=4$ respectively.  It is straightforward to use Maxwell's rule to find that the CFIP is at  $\Prob_c=2/3$, $\Prob_c=1$, respectively, at the mean-field level.  In the triangular lattice there is a second-order rigidity percolation transition and the critical point is slightly lower than $2/3$ due to critical fluctuations~\cite{Jacobs1995,Broedersz}. In the kagome case rigidity percolation is a first-order transition\cite{Mao2013c} at $\Prob=1$.  For these lattices  $\langle z \rangle = z_{\textrm{max}}\cdot (\Prob-\Prob_c)$.  These models are representative members of two  different classes of disordered networks as they approach the CFIP. As above, we make contact with the mechanics of physical biopolymer gels by identifying the lattice sites as crosslinking points and bonds as  fibrils between crosslinks. 

We write the elastic Hamiltonian for our models as~\cite{Broedersz}:
\begin{align}
	H = \sum_{\langle i,j \rangle} \frac{1}{2}k (\Delta l_{ij})^2 + \sum_{\langle i,j,k \rangle} \frac{1}{2}\kappa (\theta_{ijk})^2 .
\end{align}
The sum $\langle i,j \rangle$ runs over all bonds and $\langle i,j,k \rangle$ runs over pairs of bonds that are co-linear and share lattice site $j$.  We assign harmonic springs of spring constant $k$ for the bonds and bending stiffness $\kappa$ to the bond pairs.

In our simulations we minimize  $H$ using conjugate gradient methods to get the elastic energy, $E$. Then we numerically measured the differential shear modulus, $G =(1/A)(\partial^2E/\partial \gamma^2)$  at various values of $\Prob$ and $\kappa$. Here $A$ is the total lattice area. To shear the networks, the horizontal boundaries to which the bonds are attached are translated with strain $\gamma$. We use periodic boundary conditions  on the vertical boundaries and fixed boundary conditions  on the horizontal boundaries.  Our network size is $W=100\times 100$.  Examples of shear modulus data measured from our simulations before any scaling are shown in Fig.~\ref{FIG:lattices}cd.  It is clear from these curves that strain-stiffening is prominent when the network is diluted (smaller $\Prob$).

\begin{figure}[!ht]
	\centering
		\subfigure[]{\includegraphics[width=.4\textwidth]{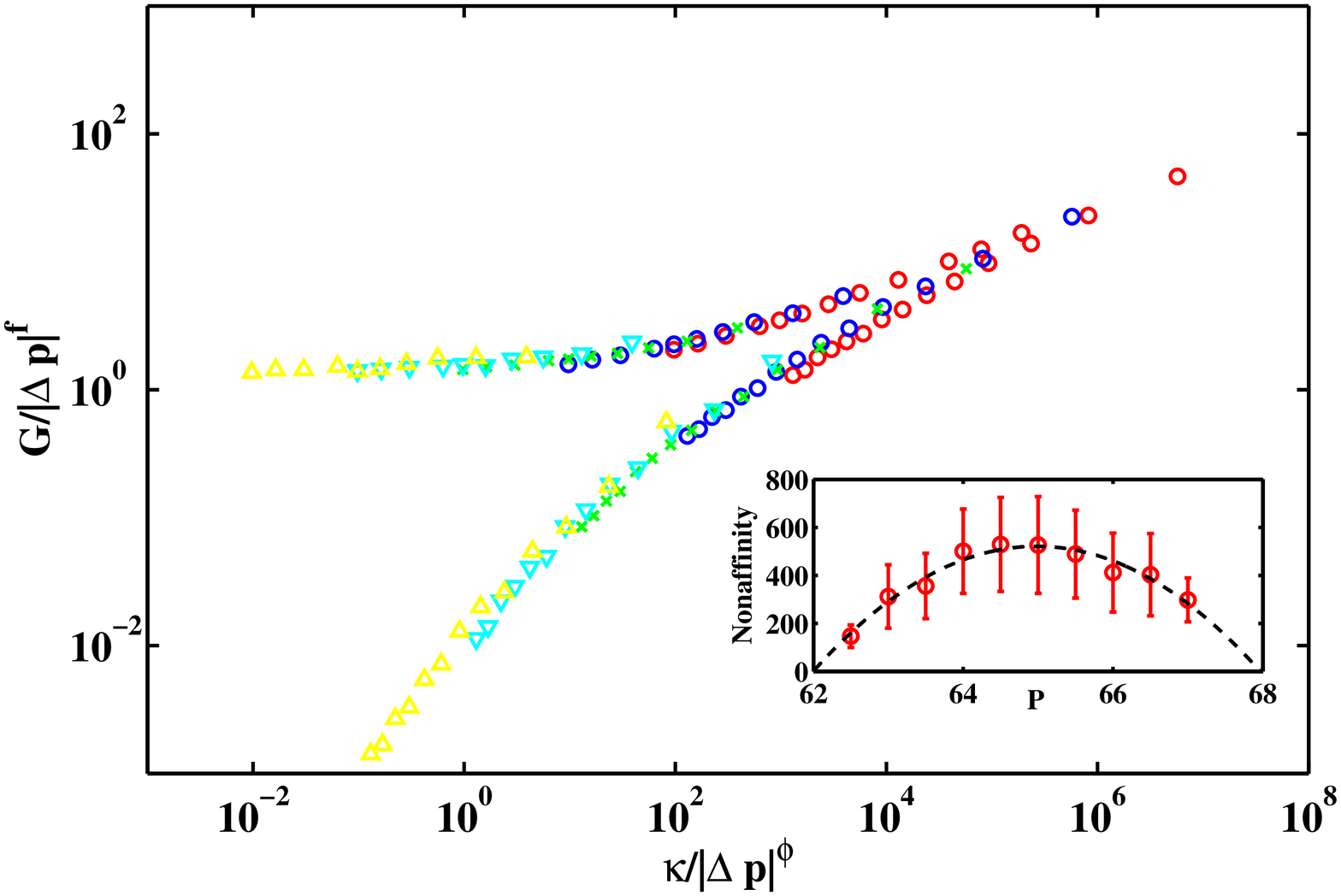}}
		\subfigure[]{\includegraphics[width=.4\textwidth]{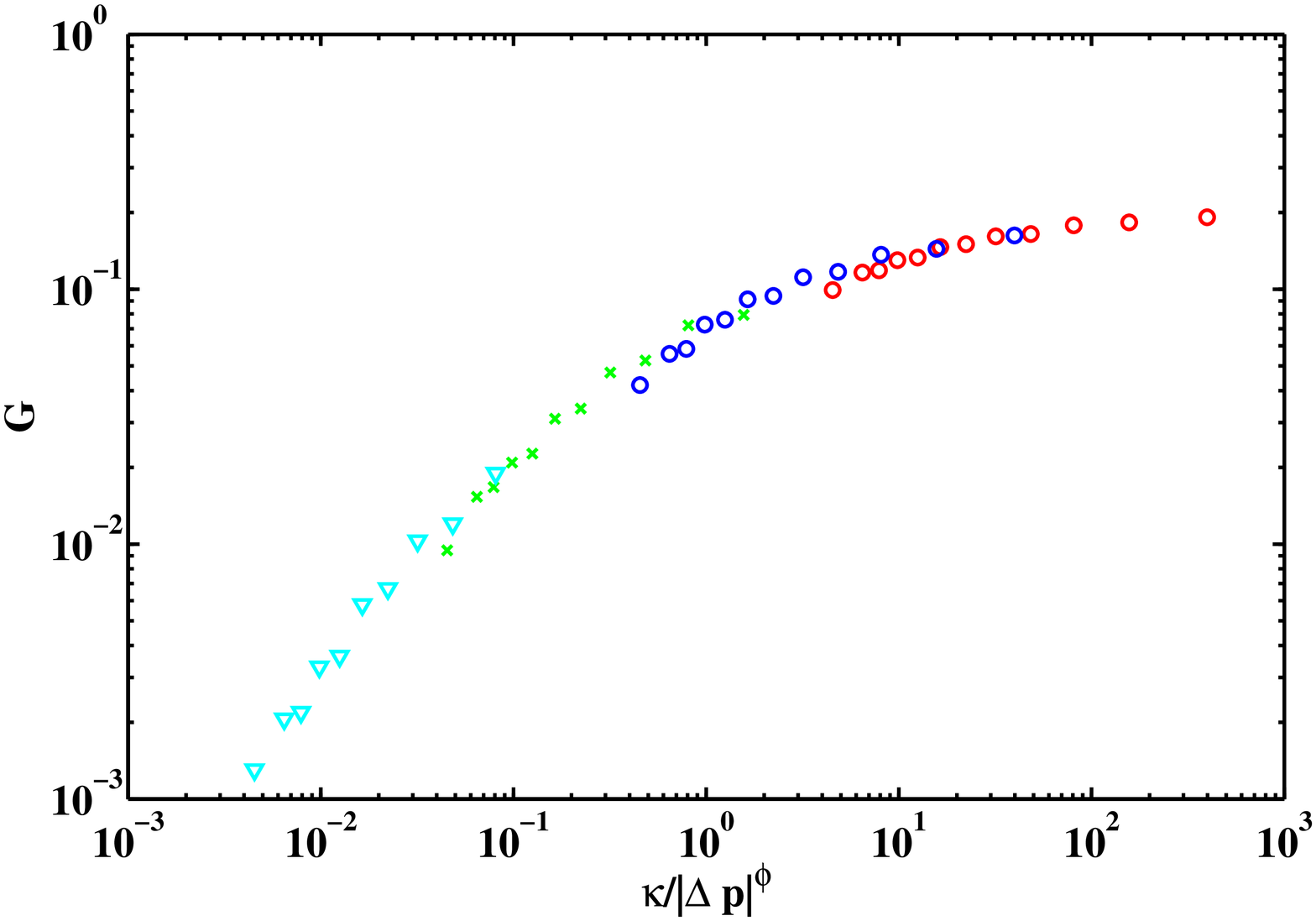}}
		\caption{\label{FIG:Linear} 
		Scaling plot for linear shear modulus $G$ (with $k=1$) at fixed $\bar \kappa$.  (a) The triangular lattice at $\bar \kappa=10^{-3}$.  (b) The kagome lattice at $\bar \kappa=10^{-4}$. The inset shows the determination of the critical point via the peak of the non-affinity parameter. }
\end{figure}
In order to test the scaling relation in Eq.~\eqref{EQ:GFull}, we need to first determine the location of the CFIP and the exponents $f$ and $\phi$.
For the triangular lattice, we follow the method of Broedersz {\it et al}\cite{Broedersz} to determine the location of the rigidity percolation point for our system size, $W$. This involves looking for a peak of the nonaffinity parameter:
\begin{align}
	\Gamma = \sum_{i} (\vec{u}_i - \vec{u}_i^{\textrm{(affine)}})^2,
\end{align}
in the central-force lattice ($\kappa=0$).  Here $i$ labels the sites, $\vec{u}_i$ is the displacement vector of the site, and $\vec{u}_i^{\textrm{(affine)}}$ is the displacement vector if the deformation would have been affine.   We find that $\Prob_c=0.651\pm 0.05$ for the triangular lattice.
Next we determine the scaling exponents $f$ and $\phi$.  These can be extracted from the small $\gamma$ (linear elasticity) limit.  As shown in Figure \ref{FIG:Linear} (a) we find that $f=1.3\pm 0.1, \phi=3.6\pm 0.3$ for the triangular lattice, consistent with previous literature on the linear elasticity of this lattice~\cite{Broedersz}. 

The CFIP for the kagome lattice is at $\Prob=1$, and we can only observe $\Prob<\Prob_c$.
Also, for this case, it is known~\cite{Mao2013c} that the transition is first order with $\Prob_c=1$; thus we expect, and indeed find $f=0$.  Our simulations show that $\phi=2.3\pm 0.2$, as shown in Fig.~\ref{FIG:Linear}b and consistent with previous literature on the linear elasticity of this lattice~\cite{Mao2013c}.

\section{Results}
\begin{figure}[!ht]
	\centering
		\subfigure[]{\includegraphics[width=.4\textwidth, height=0.26\textwidth]{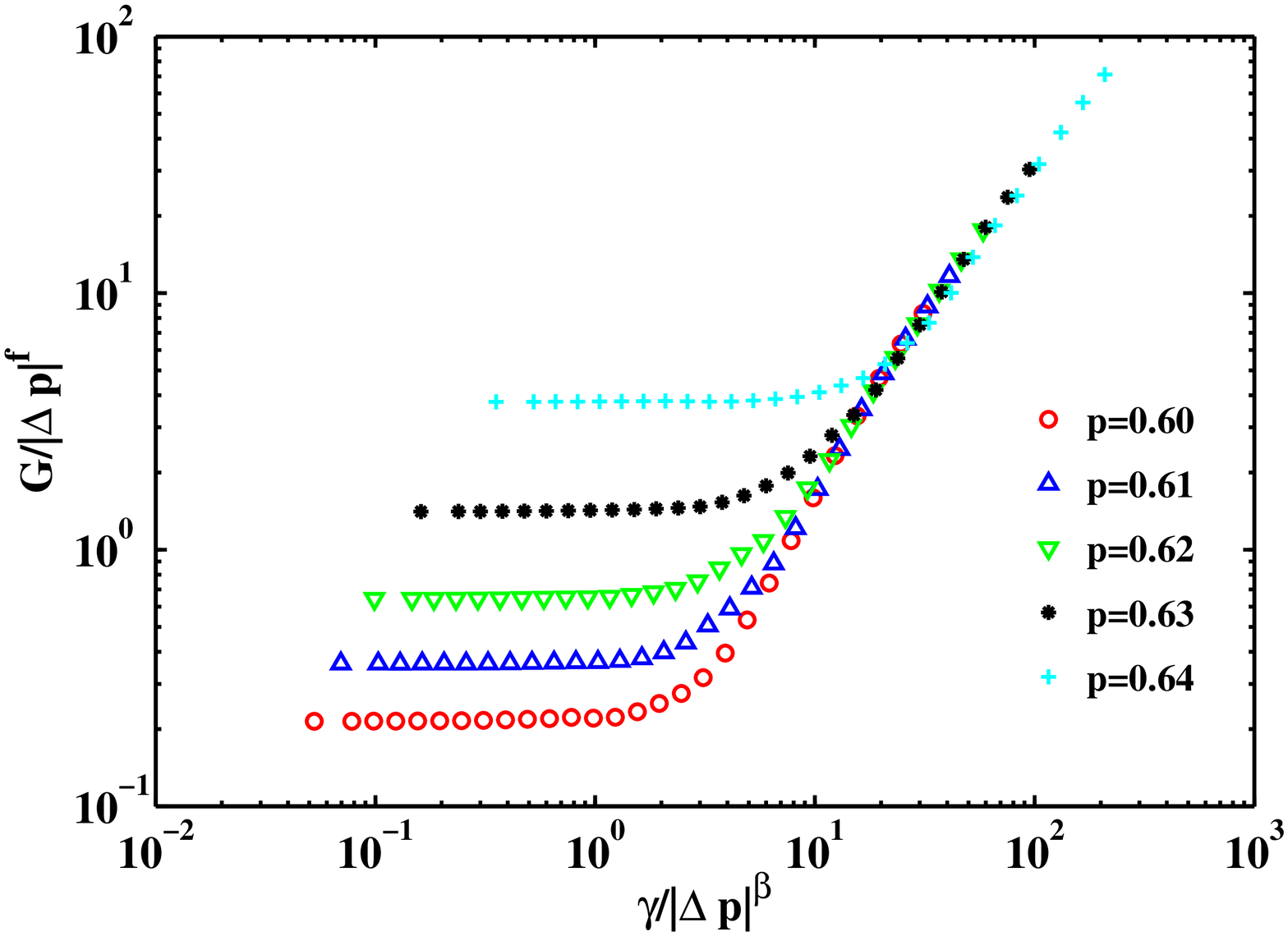}}
		\subfigure[]{\includegraphics[width=.4\textwidth, height=0.26\textwidth]{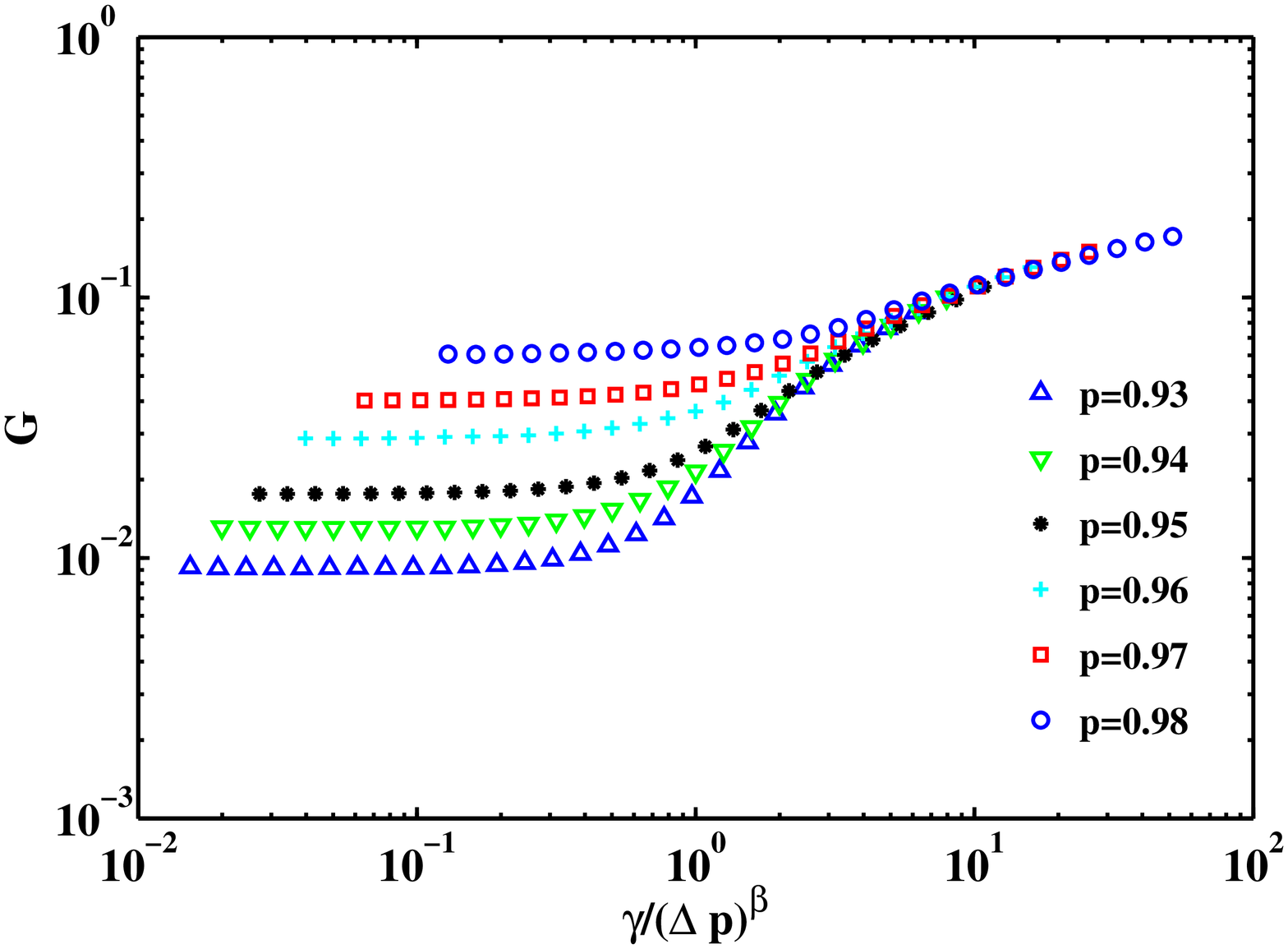}}
		\caption{\label{FIG:FixedKappa}  Scaling plot for the shear modulus at fixed $\bar \kappa$.  (a) Triangular lattice. $\kappa=10^{-3}$; Fig 2(b) Kagome lattice. $\kappa=10^{-4}$. As expected, collapse occurs only in the nonlinear regime.} 
\end{figure}

To test the scaling law of Eq.~\eqref{EQ:GFull} we can plot $G$ using scaled parameters:
\begin{equation}
	x \equiv {\bar \kappa/\vert \Delta z \vert^{\phi}}, \quad
	y \equiv {\gamma/\vert \Delta z \vert^{\beta}}.
\end{equation}
The scaling law predicts that  $G/\vert \Delta z \vert^{f}$ collapses into a \emph{single surface} as a function of $x,y$ for the correct critical exponents $f, \phi, \beta$. 
However, to verify such three-dimensional scaling collapse in general is computationally expensive.  Instead, we make two types of plots using our simulation data.  The first type of plot is at fixed $\bar \kappa$, given in Figure \ref{FIG:FixedKappa}. We obtain the expected complete collapse in the nonlinear (stretching) regime where the curves are predicted to be independent of $x$. 

\begin{figure}[!ht]
	\centering
		\subfigure[]{\includegraphics[width=.4\textwidth]{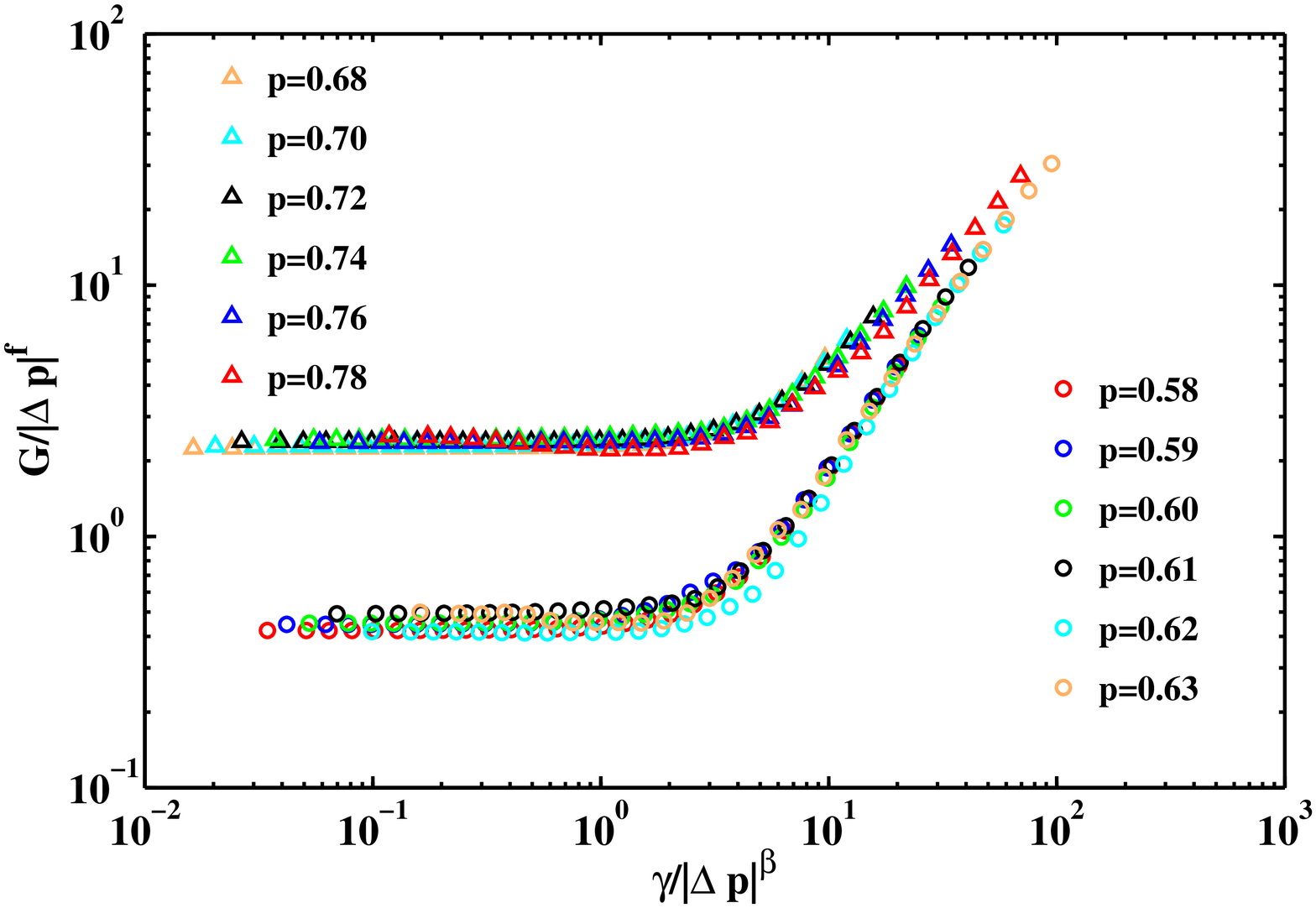}}
		\subfigure[]{\includegraphics[width=.4\textwidth]{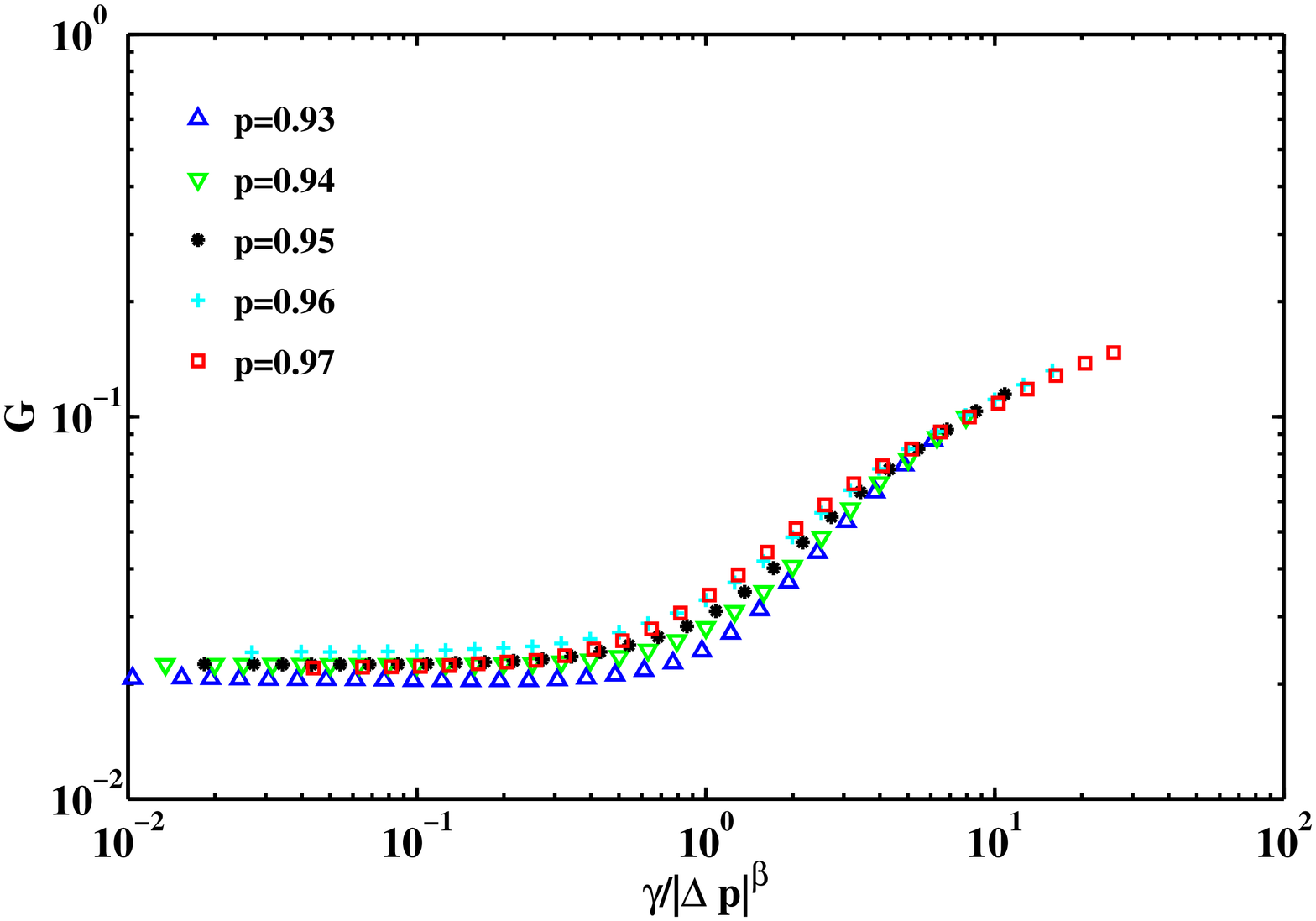}}
		\caption{ \label{FIG:Final} Two parameter scaling law Eq.~\eqref{EQ:GFull} yields good collapse in both linear and nonlinear regime when $x$ is fixed. (a) Triangular lattice, $x$=100. (b) Kagome lattice $x$=0.1}. 
\end{figure}

The second type of plot is to observe \emph{a slice} of the scaling collapse in the three-dimensional space of $x,y,G/\vert \Delta z \vert^{f}$.  To do this, we take a given finite value of $x= \bar \kappa/\vert \Delta z \vert^{\phi}$.  In practice, this correspond to taking the value of $\bar\kappa$ which changes as  $\Prob$ changes so that  $x$ is fixed in the simulation. Recall that  $\phi$ has already been determined from the linear regime.  As a result the shear modulus data completely collapses for both lattices according to our two-parameter scaling, as shown in Figure \ref{FIG:Final}.  In particular, for the triangular lattice $G(\Prob)$  lies on two branches, corresponding to $\Prob>\Prob_c$ and $\Prob<\Prob_c$.  
From these plots we find that $\beta=1.3\pm 0.1$ for the triangular lattice and $\beta=1.7\pm 0.1$ for the kagome lattice.

We should take note of a detail in the fitting procedure that we and others have used \cite{Broedersz11,Conti}. We are, in fact, never extremely close to the critical point, because sufficiently near to $p_c$ there is a non-scaling feature in the shear modulus, a dip.  We found this for both our lattice models, and it has been observed by others\cite{Broedersz11,Conti}. It occurs in models which have explicit buckling\cite{Conti} for fibrils under compression; of course, real biopolymers  buckle as well. 

The reason is easy to see. Euler buckling  is a pitchfork bifurcation as the compressive strain on an elastic beam varies. Thus we expect that the elastic energy near the buckling threshold, $\gamma_t$ will have the classic form $E = E_\circ - C(\gamma - \gamma_t)^2, \gamma > \gamma_t$. The second derivative of the energy is the shear modulus, so we expect $G \to G-C$ as we pass through the threshold. The background modulus increases, so we should see a dip in $G(\gamma)$. 

The common way~\cite{Broedersz11,Conti} to introduce buckling in model fiber networks is to introduce an extra node in the middle of each link which can bend with a small bending modulus. There is a pitchfork bifurcation at a threshold just as in the case of a continuous elastic beam.  In our models, although we do not have such buckling at the bond level, the crosslinking points represent internal degrees of freedom of the filaments and also exhibit a buckling instability.  Nevertheless, 
why these show up as a sharp dip near $p_c$ remains a puzzle to us.

\section{Conclusions and Discussion}\label{SEC:ConcDisc}
\begin{figure}[!ht]
	\centering
		\subfigure[]{\includegraphics[width=.35\textwidth]{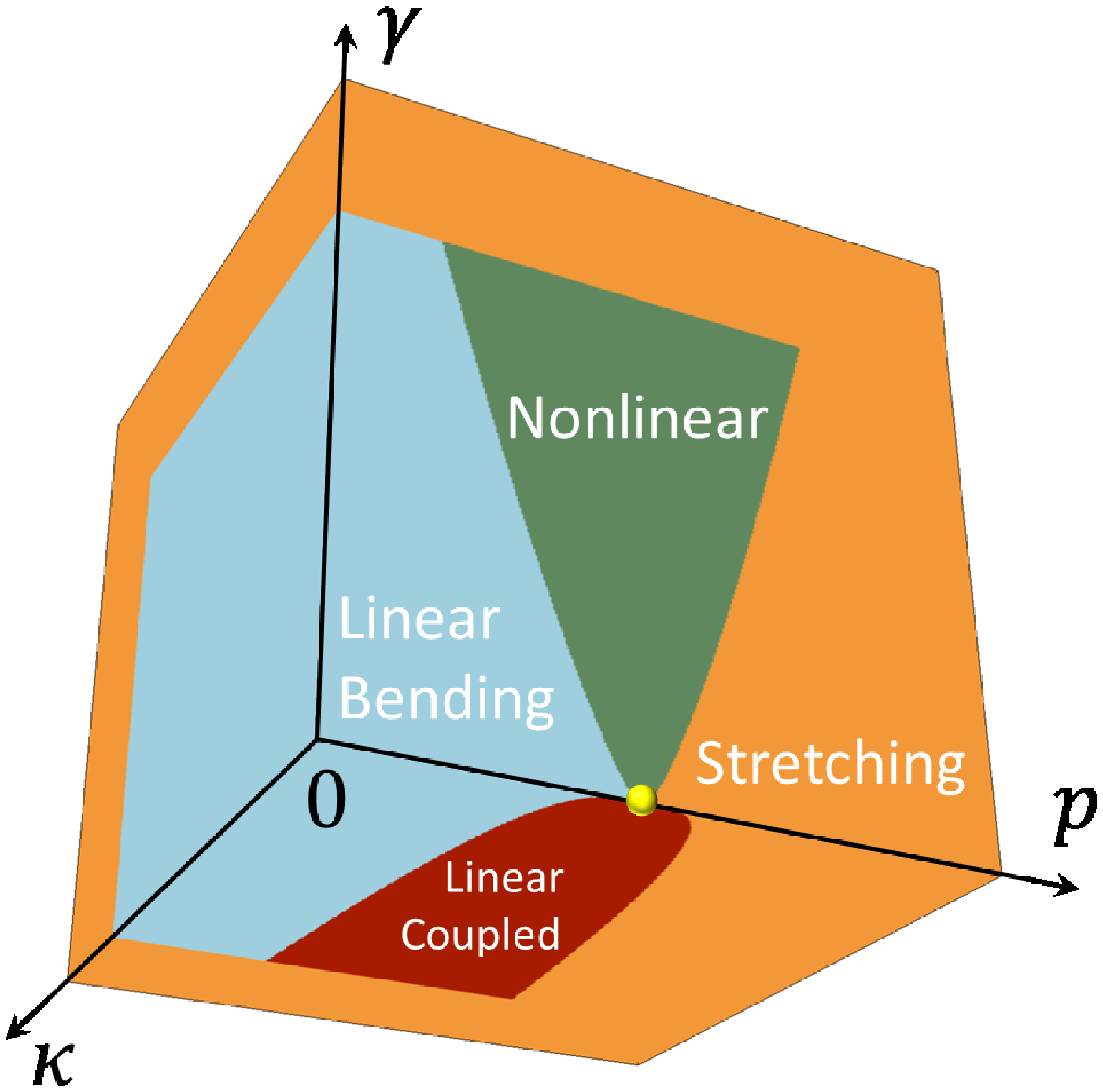}}
		\subfigure[]{\includegraphics[width=.35\textwidth]{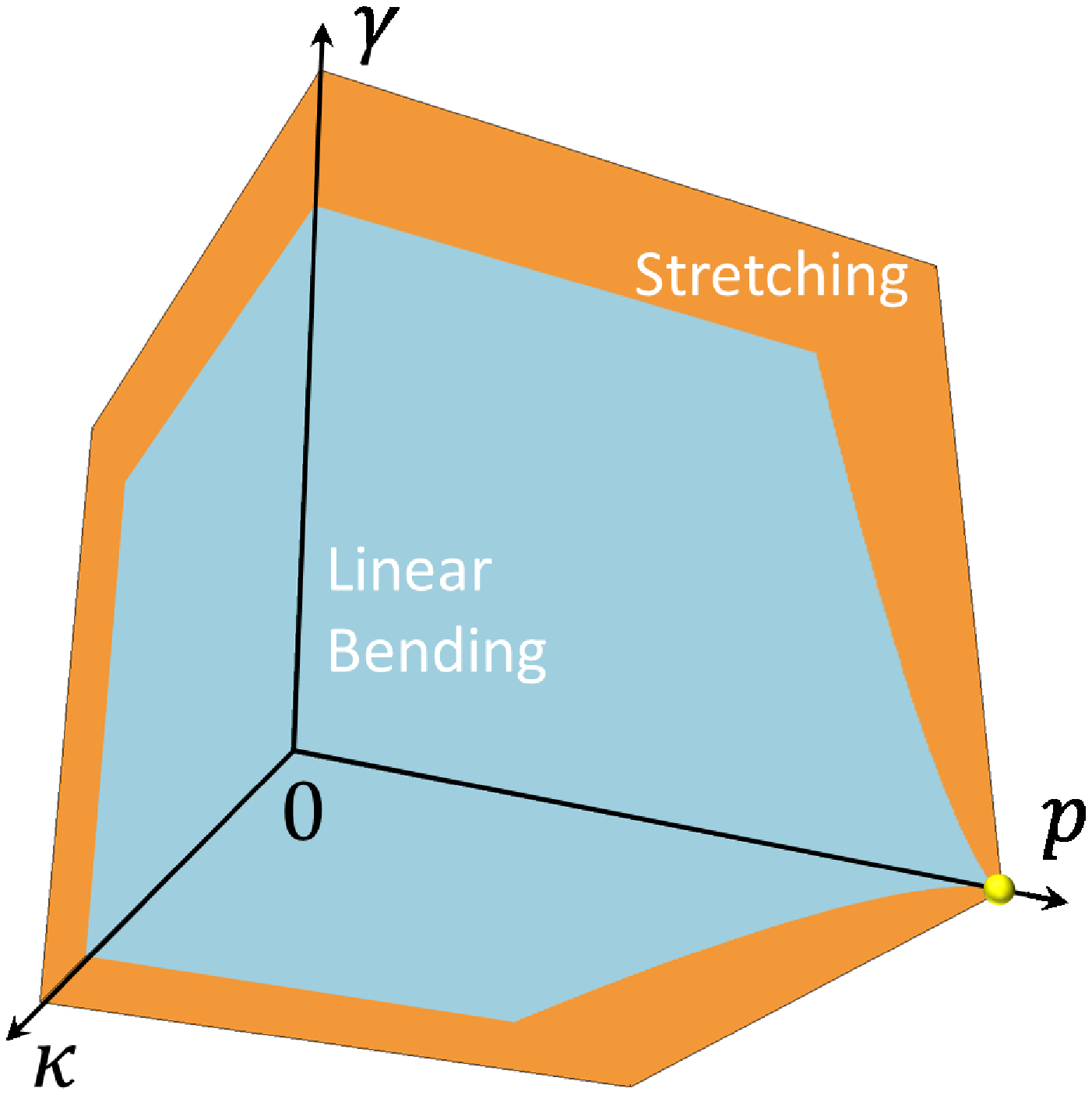}}
		\caption{ \label{FIG:PD} Three dimensional schematic phase diagrams for the triangular(a) and the kagome (b) lattices, shown in the space of dilution parameter $\Prob$, filament bending stiffness $\kappa$, and strain $\gamma$.  These phase diagrams show that strong strain-stiffening occurs when the system is bending-dominant at small deformation.} 
\end{figure}

With these results in hand, we can speculate about their applicability to other lattice models and to physical systems. Based on our results,  we can discern at least two different classes of scaling behavior with different exponents, though both obey Eq. (\ref{EQ:GFull}).  We show the schematic phase diagrams for these two classes in Fig.~\ref{FIG:PD}.

In the first  case, exemplified by the diluted triangular lattice, we have a network that can vary through the CFIP as a parameter is changed. Biopolymer gels with $z_{\textrm{max}}>2d$ belong to this class.  Some dense forms of collagen, such as tendons and cartilage may also belong to this class.  
In this case  $z_{\textrm{max}}>2d$ so that we can tune through $\Prob_c<1$. Then we have various regimes implied by Eq. (\ref{EQ:GFull}):
\begin{itemize}
	\item A linear bending regime when $\Delta z<0, x\ll 1, y\ll 1$.  This regime is described by linear elasticity so $\mathcal{G}_{-}$ is independent of $y$. This regime is bending dominated: 
	\begin{equation}
		\mathcal{G}_{-}(x,y) \sim x \quad		G \sim \kappa |\Delta z|^{f-\phi} .
	\end{equation} 
	\item A linear coupled regime when $\Delta z<0, x\gg 1, y\ll 1$ while  $\bar \kappa \ll 1$.  This regime is also described by linear elasticity; $\mathcal{G}_{-}$ is independent of $y$.  Also,  it is in a critical regime ($x \gg 1$) so $G$ should be independent of $\Delta z$.  Thus:
	\begin{equation}
		\mathcal{G}_{-}(x,y) \sim x^{f/\phi} \quad
		G \sim \kappa^{f/\phi} k^{1-f/\phi}
	\end{equation}
	\item A nonlinear regime when $y\gg 1$ while keeping $\gamma\ll 1$.  This regime characterizes nonlinear elasticity as the strain goes beyond a ``turning point''  $\gamma^*\sim |\Delta z|^{\beta}$.  The turning point approaches $0$ as  to $\Prob \to \Prob_c$ meaning that near the CFIP the linear elasticity regime vanishes.  It is reasonable to hypothesize that in this regime $G$ is independent of $\Delta z$ and $x$ so that:
	\begin{equation}
		\mathcal{G}_{\pm}(x,y)  \sim y^{f/\beta} \quad
		G \sim k \gamma^{f/\beta} .
	\end{equation} 
	Our simulation data verifies this hypothesis.
	\item The stretching regime.  This includes a small-strain stretching-dominated regime when $\Delta z>0$ and $\bar \kappa\ll 1$, and a large-strain stretching-dominant regime when $\gamma\gg 1$ and the bending modes are all stretched out.  Here the elasticity is controlled by the stretching stiffness $k$:
	\begin{equation}
		\mathcal{G}_{\pm}(x,y) \sim 1 \quad
		G \sim k |\Delta z|^{f} .
	\end{equation} 
\end{itemize}

The other category, for which the diluted kagome lattice is an example, has $z_{\textrm{max}}=2d$ and a first-order central-force rigidity transition at $\Prob_c=1$ and $f=0$.  In this case we have two regimes: 
\begin{itemize}
	\item A  linear bending regime when $x\ll 1, y\ll 1$ (in this case we always have $\Delta z<0$) so that:
	\begin{align}
		\mathcal{G}_{-}(x,y) &\sim x \quad
		G \sim \kappa |\Delta z|^{-\phi} .
	\end{align} 
	\item The stretching regime, including small-strain part $x \gg 1$ and large-strain part $\gamma \gg 1$.  In this regime deformations are dominated by stretching of fibers and
	\begin{align}
		\mathcal{G}_{\pm}(x,y) \sim 1 \quad
		G \sim k  .
	\end{align} 
\end{itemize}
The linear coupled regime and the nonlinear regime are missing  because they are consequences of a continuous transition at $\Prob_c<1$.  As a result the crossover to stretching  as strain increases is more abrupt in these lattices.

Many three-dimensional fiber-networks including dilute collagen I, have $z_{\textrm{max}}<2d$.  For this case, the central-force network is not rigid even at $\Prob=1$.  As in the kagome case, scaling law predicts a linear bending regime crossing over to stretching  at either large $\kappa$ or large $\gamma$.  This is consistent with numerical results on three-dimensional models~\cite{Broedersz,Stenull2011,Broedersz2012,Huisman2011}. 

We can relate experimental results to our models by focusing on three dimensionless numbers. The first one is the maximum coordination of the crosslinkers $z_{\textrm{max}}$ which determines which type of the phase diagrams applies.  The second one is $\bar \kappa$ which can be deduced, for example, from the Young's modulus and the shape of individual fibers. The simplest case of this estimate has been given above. The third one is $\Prob$. For a network made up of fibers of mean length $L$ it is easy to see that the relation between mesh size $a$ and fiber length $L$ is:
$a = L/(1-\Prob)$. This allows us to estimate the effective bond dilution from data.  With these three numbers one can sketch the phase diagram and determine which regime the experimental system belongs to.

In addition, it is also interesting to compare our theory to other types of gels with less stiff filaments and thus stronger thermal fluctuations.  A different mechanism of strain-stiffening has been proposed in the literature attributing strain stiffening to stretching out thermal undulations of polymers~\cite{MacKintosh1995,Storm2005}.  This is relevant for gels composed of filaments such as actin or elastin which have persistence length comparable or smaller than the mesh size, so considerable thermal undulations present, thus providing a regime of entropic elasticity at small strain.  Although for these gels stretching out thermal undulations contribute to the strain-stiffening, if the gel is below the CFIP (which is true for most bio-polymer gels), the bending modes still need to be stretched out in addition to the thermal undulations before the gel enters true stretching-dominant elasticity regime.

During the preparation of our manuscript, we learned about two recent preprints~\cite{Licup2015,Sharma} which use a combination of experimental and simulation tools to investigate nonlinear elasticity.  Both of these preprints mainly focus on the nonlinear shear modulus as bending stiffness $\kappa$ varies while keeping the connectivity $\langle z \rangle$ fixed.  In contrast, our study obtained full collapse of shear modulus in the more general case of varying both $\kappa$ and $\langle z \rangle$.  Moreover, we also identify various regimes of nonlinear elasticity in a three-dimensional phase diagram of connectivity, bending stiffness, and strain.

\section*{Acknowledgments}  We acknowledge informative discussions with F. C. MacKintosh, C. P. Broedersz.  
J-C Feng is supported by the National Science Foundation Center for Theoretical Biological Physics (Grant PHY-1427654). The work of HL was supported in part by the Cancer Prevention and Research Institute of Texas
(CPRIT) Scholar Program of the State of Texas at Rice University.


\begin{thebibliography}{33}
\expandafter\ifx\csname natexlab\endcsname\relax\def\natexlab#1{#1}\fi
\expandafter\ifx\csname bibnamefont\endcsname\relax
  \def\bibnamefont#1{#1}\fi
\expandafter\ifx\csname bibfnamefont\endcsname\relax
  \def\bibfnamefont#1{#1}\fi
\expandafter\ifx\csname citenamefont\endcsname\relax
  \def\citenamefont#1{#1}\fi
\expandafter\ifx\csname url\endcsname\relax
  \def\url#1{\texttt{#1}}\fi
\expandafter\ifx\csname urlprefix\endcsname\relax\def\urlprefix{URL }\fi
\providecommand{\bibinfo}[2]{#2}
\providecommand{\eprint}[2][]{\url{#2}}

\bibitem[{\citenamefont{Roeder et~al.}(2002)\citenamefont{Roeder, Kokini,
  Sturgis, Robinson, and Voytik-Harbin}}]{Roeder}
\bibinfo{author}{\bibfnamefont{B.}~\bibnamefont{Roeder}},
  \bibinfo{author}{\bibfnamefont{K.}~\bibnamefont{Kokini}},
  \bibinfo{author}{\bibfnamefont{J.}~\bibnamefont{Sturgis}},
  \bibinfo{author}{\bibfnamefont{J.}~\bibnamefont{Robinson}}, \bibnamefont{and}
  \bibinfo{author}{\bibfnamefont{S.}~\bibnamefont{Voytik-Harbin}},
  \bibinfo{journal}{Journal of biomechanical engineering}
  \textbf{\bibinfo{volume}{124}}, \bibinfo{pages}{214} (\bibinfo{year}{2002}).

\bibitem[{\citenamefont{Stein et~al.}(2011)\citenamefont{Stein, Vader, Weitz,
  and Sander}}]{Stein}
\bibinfo{author}{\bibfnamefont{A.~M.} \bibnamefont{Stein}},
  \bibinfo{author}{\bibfnamefont{D.~A.} \bibnamefont{Vader}},
  \bibinfo{author}{\bibfnamefont{D.~A.} \bibnamefont{Weitz}}, \bibnamefont{and}
  \bibinfo{author}{\bibfnamefont{L.~M.} \bibnamefont{Sander}},
  \bibinfo{journal}{Complexity} \textbf{\bibinfo{volume}{16}},
  \bibinfo{pages}{22} (\bibinfo{year}{2011}).

\bibitem[{\citenamefont{Storm et~al.}(2005)\citenamefont{Storm, Pastore,
  MacKintosh, Lubensky, and Janmey}}]{Storm2005}
\bibinfo{author}{\bibfnamefont{C.}~\bibnamefont{Storm}},
  \bibinfo{author}{\bibfnamefont{J.}~\bibnamefont{Pastore}},
  \bibinfo{author}{\bibfnamefont{F.}~\bibnamefont{MacKintosh}},
  \bibinfo{author}{\bibfnamefont{T.}~\bibnamefont{Lubensky}}, \bibnamefont{and}
  \bibinfo{author}{\bibfnamefont{P.}~\bibnamefont{Janmey}},
  \bibinfo{journal}{Nature} \textbf{\bibinfo{volume}{435}},
  \bibinfo{pages}{191} (\bibinfo{year}{2005}).

\bibitem[{\citenamefont{Broedersz et~al.}(2011)\citenamefont{Broedersz, Mao,
  Lubensky, and MacKintosh}}]{Broedersz}
\bibinfo{author}{\bibfnamefont{C.~P.} \bibnamefont{Broedersz}},
  \bibinfo{author}{\bibfnamefont{X.}~\bibnamefont{Mao}},
  \bibinfo{author}{\bibfnamefont{T.~C.} \bibnamefont{Lubensky}},
  \bibnamefont{and} \bibinfo{author}{\bibfnamefont{F.~C.}
  \bibnamefont{MacKintosh}}, \bibinfo{journal}{Nature Physics}
  \textbf{\bibinfo{volume}{7}}, \bibinfo{pages}{983} (\bibinfo{year}{2011}).

\bibitem[{\citenamefont{Broedersz and MacKintosh}(2014)}]{Broedersz2014}
\bibinfo{author}{\bibfnamefont{C.~P.} \bibnamefont{Broedersz}}
  \bibnamefont{and} \bibinfo{author}{\bibfnamefont{F.~C.}
  \bibnamefont{MacKintosh}}, \bibinfo{journal}{Rev. Mod. Phys.}
  \textbf{\bibinfo{volume}{86}}, \bibinfo{pages}{995} (\bibinfo{year}{2014}).

\bibitem[{\citenamefont{Provenzano et~al.}(2006)\citenamefont{Provenzano,
  Eliceiri, Campbell, Inman, White, and Keely}}]{Provenzano}
\bibinfo{author}{\bibfnamefont{P.}~\bibnamefont{Provenzano}},
  \bibinfo{author}{\bibfnamefont{K.}~\bibnamefont{Eliceiri}},
  \bibinfo{author}{\bibfnamefont{J.}~\bibnamefont{Campbell}},
  \bibinfo{author}{\bibfnamefont{D.}~\bibnamefont{Inman}},
  \bibinfo{author}{\bibfnamefont{J.}~\bibnamefont{White}}, \bibnamefont{and}
  \bibinfo{author}{\bibfnamefont{P.}~\bibnamefont{Keely}},
  \bibinfo{journal}{BMC medicine} \textbf{\bibinfo{volume}{4}},
  \bibinfo{pages}{38} (\bibinfo{year}{2006}).

\bibitem[{\citenamefont{Onck et~al.}(2005)\citenamefont{Onck, Koeman, van
  Dillen, and van~der Giessen}}]{Onck}
\bibinfo{author}{\bibfnamefont{P.~R.} \bibnamefont{Onck}},
  \bibinfo{author}{\bibfnamefont{T.}~\bibnamefont{Koeman}},
  \bibinfo{author}{\bibfnamefont{T.}~\bibnamefont{van Dillen}},
  \bibnamefont{and} \bibinfo{author}{\bibfnamefont{E.}~\bibnamefont{van~der
  Giessen}}, \bibinfo{journal}{Phys. Rev. Lett.} \textbf{\bibinfo{volume}{95}},
  \bibinfo{pages}{178102} (\bibinfo{year}{2005}).

\bibitem[{\citenamefont{Wyart et~al.}(2008)\citenamefont{Wyart, Liang, Kabla,
  and Mahadevan}}]{Wyart}
\bibinfo{author}{\bibfnamefont{M.}~\bibnamefont{Wyart}},
  \bibinfo{author}{\bibfnamefont{H.}~\bibnamefont{Liang}},
  \bibinfo{author}{\bibfnamefont{A.}~\bibnamefont{Kabla}}, \bibnamefont{and}
  \bibinfo{author}{\bibfnamefont{L.}~\bibnamefont{Mahadevan}},
  \bibinfo{journal}{Phys. Rev. Lett.} \textbf{\bibinfo{volume}{101}},
  \bibinfo{pages}{215501} (\bibinfo{year}{2008}).

\bibitem[{\citenamefont{Sheinman et~al.}(2012)\citenamefont{Sheinman,
  Broedersz, and MacKintosh}}]{Sheinman2012}
\bibinfo{author}{\bibfnamefont{M.}~\bibnamefont{Sheinman}},
  \bibinfo{author}{\bibfnamefont{C.~P.} \bibnamefont{Broedersz}},
  \bibnamefont{and} \bibinfo{author}{\bibfnamefont{F.~C.}
  \bibnamefont{MacKintosh}}, \bibinfo{journal}{Phys. Rev. E}
  \textbf{\bibinfo{volume}{85}}, \bibinfo{pages}{021801}
  (\bibinfo{year}{2012}).

\bibitem[{\citenamefont{Feng et~al.}(2015)\citenamefont{Feng, Levine, Mao, and
  Sander}}]{Feng2015}
\bibinfo{author}{\bibfnamefont{J.}~\bibnamefont{Feng}},
  \bibinfo{author}{\bibfnamefont{H.}~\bibnamefont{Levine}},
  \bibinfo{author}{\bibfnamefont{X.}~\bibnamefont{Mao}}, \bibnamefont{and}
  \bibinfo{author}{\bibfnamefont{L.~M.} \bibnamefont{Sander}},
  \bibinfo{journal}{Phys. Rev. E} \textbf{\bibinfo{volume}{91}},
  \bibinfo{pages}{042710} (\bibinfo{year}{2015}).

\bibitem[{\citenamefont{Alexander}(1998)}]{Alexander1998}
\bibinfo{author}{\bibfnamefont{S.}~\bibnamefont{Alexander}},
  \bibinfo{journal}{Phys. Rep.} \textbf{\bibinfo{volume}{296}},
  \bibinfo{pages}{65} (\bibinfo{year}{1998}).

\bibitem[{\citenamefont{Maxwell}(1864)}]{Maxwell1864}
\bibinfo{author}{\bibfnamefont{J.~C.} \bibnamefont{Maxwell}},
  \bibinfo{journal}{Philos. Mag.} \textbf{\bibinfo{volume}{27}},
  \bibinfo{pages}{294} (\bibinfo{year}{1864}).

\bibitem[{\citenamefont{Liu et~al.}(2010)\citenamefont{Liu, Nagel, van
  Saarloos, and Wyart}}]{Liu2010}
\bibinfo{author}{\bibfnamefont{A.~J.} \bibnamefont{Liu}},
  \bibinfo{author}{\bibfnamefont{S.~R.} \bibnamefont{Nagel}},
  \bibinfo{author}{\bibfnamefont{W.}~\bibnamefont{van Saarloos}},
  \bibnamefont{and} \bibinfo{author}{\bibfnamefont{M.}~\bibnamefont{Wyart}}, in
  \emph{\bibinfo{booktitle}{Dynamical heterogeneities in glasses, colloids, and
  granular media}}, edited by
  \bibinfo{editor}{\bibfnamefont{L.}~\bibnamefont{Berthier}},
  \bibinfo{editor}{\bibfnamefont{G.}~\bibnamefont{Biroli}},
  \bibinfo{editor}{\bibfnamefont{J.-P.} \bibnamefont{Bouchaud}},
  \bibinfo{editor}{\bibfnamefont{L.}~\bibnamefont{Cipeletti}},
  \bibnamefont{and} \bibinfo{editor}{\bibfnamefont{W.}~\bibnamefont{van
  Saarloos}} (\bibinfo{publisher}{Oxford University Press},
  \bibinfo{year}{2010}), chap.~\bibinfo{chapter}{9}.

\bibitem[{\citenamefont{Mao et~al.}(2010)\citenamefont{Mao, Xu, and
  Lubensky}}]{Mao2010}
\bibinfo{author}{\bibfnamefont{X.}~\bibnamefont{Mao}},
  \bibinfo{author}{\bibfnamefont{N.}~\bibnamefont{Xu}}, \bibnamefont{and}
  \bibinfo{author}{\bibfnamefont{T.~C.} \bibnamefont{Lubensky}},
  \bibinfo{journal}{Phys. Rev. Lett.} \textbf{\bibinfo{volume}{104}},
  \bibinfo{pages}{085504} (\bibinfo{year}{2010}).

\bibitem[{\citenamefont{Mao and Lubensky}(2011)}]{Mao}
\bibinfo{author}{\bibfnamefont{X.}~\bibnamefont{Mao}} \bibnamefont{and}
  \bibinfo{author}{\bibfnamefont{T.~C.} \bibnamefont{Lubensky}},
  \bibinfo{journal}{Phys. Rev. E} \textbf{\bibinfo{volume}{83}},
  \bibinfo{pages}{011111} (\bibinfo{year}{2011}).

\bibitem[{\citenamefont{Ellenbroek and Mao}(2011)}]{Ellenbroek2011}
\bibinfo{author}{\bibfnamefont{W.~G.} \bibnamefont{Ellenbroek}}
  \bibnamefont{and} \bibinfo{author}{\bibfnamefont{X.}~\bibnamefont{Mao}},
  \bibinfo{journal}{Europhys. Lett.} \textbf{\bibinfo{volume}{96}},
  \bibinfo{pages}{52002} (\bibinfo{year}{2011}).

\bibitem[{\citenamefont{Mao et~al.}(2015)\citenamefont{Mao, Souslov, Mendoza,
  and Lubensky}}]{Mao2015}
\bibinfo{author}{\bibfnamefont{X.}~\bibnamefont{Mao}},
  \bibinfo{author}{\bibfnamefont{A.}~\bibnamefont{Souslov}},
  \bibinfo{author}{\bibfnamefont{C.~I.} \bibnamefont{Mendoza}},
  \bibnamefont{and} \bibinfo{author}{\bibfnamefont{T.~C.}
  \bibnamefont{Lubensky}}, \bibinfo{journal}{Nat. Commun.}
  \textbf{\bibinfo{volume}{6}}, \bibinfo{pages}{5968} (\bibinfo{year}{2015}).

\bibitem[{\citenamefont{Rocklin and Mao}(2014)}]{Rocklin2014}
\bibinfo{author}{\bibfnamefont{D.~Z.} \bibnamefont{Rocklin}} \bibnamefont{and}
  \bibinfo{author}{\bibfnamefont{X.}~\bibnamefont{Mao}}, \bibinfo{journal}{Soft
  Matter} \textbf{\bibinfo{volume}{10}}, \bibinfo{pages}{7569}
  (\bibinfo{year}{2014}).

\bibitem[{\citenamefont{Zhang et~al.}(2015)\citenamefont{Zhang, Rocklin, Chen,
  and Mao}}]{Zhang2015a}
\bibinfo{author}{\bibfnamefont{L.}~\bibnamefont{Zhang}},
  \bibinfo{author}{\bibfnamefont{D.~Z.} \bibnamefont{Rocklin}},
  \bibinfo{author}{\bibfnamefont{B.~G.-g.} \bibnamefont{Chen}},
  \bibnamefont{and} \bibinfo{author}{\bibfnamefont{X.}~\bibnamefont{Mao}},
  \bibinfo{journal}{Phys. Rev. E} \textbf{\bibinfo{volume}{91}},
  \bibinfo{pages}{032124} (\bibinfo{year}{2015}).

\bibitem[{\citenamefont{Lubensky et~al.}(2015)\citenamefont{Lubensky, Kane,
  Mao, Souslov, and Sun}}]{Lubensky2015}
\bibinfo{author}{\bibfnamefont{T.~C.} \bibnamefont{Lubensky}},
  \bibinfo{author}{\bibfnamefont{C.}~\bibnamefont{Kane}},
  \bibinfo{author}{\bibfnamefont{X.}~\bibnamefont{Mao}},
  \bibinfo{author}{\bibfnamefont{A.}~\bibnamefont{Souslov}}, \bibnamefont{and}
  \bibinfo{author}{\bibfnamefont{K.}~\bibnamefont{Sun}},
  \bibinfo{journal}{Reports on Progress in Physics}
  \textbf{\bibinfo{volume}{78}}, \bibinfo{pages}{073901}
  (\bibinfo{year}{2015}).

\bibitem[{\citenamefont{Mao et~al.}(2013{\natexlab{a}})\citenamefont{Mao,
  Stenull, and Lubensky}}]{Mao2013b}
\bibinfo{author}{\bibfnamefont{X.}~\bibnamefont{Mao}},
  \bibinfo{author}{\bibfnamefont{O.}~\bibnamefont{Stenull}}, \bibnamefont{and}
  \bibinfo{author}{\bibfnamefont{T.~C.} \bibnamefont{Lubensky}},
  \bibinfo{journal}{Phys. Rev. E} \textbf{\bibinfo{volume}{87}},
  \bibinfo{pages}{042601} (\bibinfo{year}{2013}{\natexlab{a}}).

\bibitem[{\citenamefont{Mao et~al.}(2013{\natexlab{b}})\citenamefont{Mao,
  Stenull, and Lubensky}}]{Mao2013c}
\bibinfo{author}{\bibfnamefont{X.}~\bibnamefont{Mao}},
  \bibinfo{author}{\bibfnamefont{O.}~\bibnamefont{Stenull}}, \bibnamefont{and}
  \bibinfo{author}{\bibfnamefont{T.~C.} \bibnamefont{Lubensky}},
  \bibinfo{journal}{Phys. Rev. E} \textbf{\bibinfo{volume}{87}},
  \bibinfo{pages}{042602} (\bibinfo{year}{2013}{\natexlab{b}}).

\bibitem[{\citenamefont{Das et~al.}(2007)\citenamefont{Das, MacKintosh, and
  Levine}}]{Das2007}
\bibinfo{author}{\bibfnamefont{M.}~\bibnamefont{Das}},
  \bibinfo{author}{\bibfnamefont{F.~C.} \bibnamefont{MacKintosh}},
  \bibnamefont{and} \bibinfo{author}{\bibfnamefont{A.~J.}
  \bibnamefont{Levine}}, \bibinfo{journal}{Phys. Rev. Lett.}
  \textbf{\bibinfo{volume}{99}}, \bibinfo{pages}{038101}
  (\bibinfo{year}{2007}).

\bibitem[{\citenamefont{Das et~al.}(2012)\citenamefont{Das, Quint, and
  Schwarz}}]{Das2012}
\bibinfo{author}{\bibfnamefont{M.}~\bibnamefont{Das}},
  \bibinfo{author}{\bibfnamefont{D.~A.} \bibnamefont{Quint}}, \bibnamefont{and}
  \bibinfo{author}{\bibfnamefont{J.~M.} \bibnamefont{Schwarz}},
  \bibinfo{journal}{PloS one} \textbf{\bibinfo{volume}{7}},
  \bibinfo{pages}{e35939} (\bibinfo{year}{2012}).

\bibitem[{\citenamefont{Jacobs and Thorpe}(1995)}]{Jacobs1995}
\bibinfo{author}{\bibfnamefont{D.~J.} \bibnamefont{Jacobs}} \bibnamefont{and}
  \bibinfo{author}{\bibfnamefont{M.~F.} \bibnamefont{Thorpe}},
  \bibinfo{journal}{Phys. Rev. Lett.} \textbf{\bibinfo{volume}{75}},
  \bibinfo{pages}{4051} (\bibinfo{year}{1995}).

\bibitem[{\citenamefont{Broedersz and MacKintosh}(2011)}]{Broedersz11}
\bibinfo{author}{\bibfnamefont{C.~P.} \bibnamefont{Broedersz}}
  \bibnamefont{and} \bibinfo{author}{\bibfnamefont{F.~C.}
  \bibnamefont{MacKintosh}}, \bibinfo{journal}{Soft Matter}
  \textbf{\bibinfo{volume}{7}}, \bibinfo{pages}{3186} (\bibinfo{year}{2011}).

\bibitem[{\citenamefont{Conti and Mackintosh}(2009)}]{Conti}
\bibinfo{author}{\bibfnamefont{E.}~\bibnamefont{Conti}} \bibnamefont{and}
  \bibinfo{author}{\bibfnamefont{F.~C.} \bibnamefont{Mackintosh}},
  \bibinfo{journal}{Physical Review Letters} \textbf{\bibinfo{volume}{102}},
  \bibinfo{pages}{088102} (\bibinfo{year}{2009}).

\bibitem[{\citenamefont{Stenull and Lubensky}(2011)}]{Stenull2011}
\bibinfo{author}{\bibfnamefont{O.}~\bibnamefont{Stenull}} \bibnamefont{and}
  \bibinfo{author}{\bibfnamefont{T.}~\bibnamefont{Lubensky}},
  \bibinfo{journal}{arXiv preprint arXiv:1108.4328}  (\bibinfo{year}{2011}).

\bibitem[{\citenamefont{Broedersz et~al.}(2012)\citenamefont{Broedersz,
  Sheinman, and MacKintosh}}]{Broedersz2012}
\bibinfo{author}{\bibfnamefont{C.~P.} \bibnamefont{Broedersz}},
  \bibinfo{author}{\bibfnamefont{M.}~\bibnamefont{Sheinman}}, \bibnamefont{and}
  \bibinfo{author}{\bibfnamefont{F.~C.} \bibnamefont{MacKintosh}},
  \bibinfo{journal}{Phys. Rev. Lett.} \textbf{\bibinfo{volume}{108}},
  \bibinfo{pages}{078102} (\bibinfo{year}{2012}).

\bibitem[{\citenamefont{Huisman and Lubensky}(2011)}]{Huisman2011}
\bibinfo{author}{\bibfnamefont{E.~M.} \bibnamefont{Huisman}} \bibnamefont{and}
  \bibinfo{author}{\bibfnamefont{T.~C.} \bibnamefont{Lubensky}},
  \bibinfo{journal}{Phys. Rev. Lett.} \textbf{\bibinfo{volume}{106}},
  \bibinfo{pages}{088301} (\bibinfo{year}{2011}).

\bibitem[{\citenamefont{MacKintosh et~al.}(1995)\citenamefont{MacKintosh,
  K\"as, and Janmey}}]{MacKintosh1995}
\bibinfo{author}{\bibfnamefont{F.~C.} \bibnamefont{MacKintosh}},
  \bibinfo{author}{\bibfnamefont{J.}~\bibnamefont{K\"as}}, \bibnamefont{and}
  \bibinfo{author}{\bibfnamefont{P.~A.} \bibnamefont{Janmey}},
  \bibinfo{journal}{Phys. Rev. Lett.} \textbf{\bibinfo{volume}{75}},
  \bibinfo{pages}{4425} (\bibinfo{year}{1995}).

\bibitem[{\citenamefont{Licup et~al.}(2015)\citenamefont{Licup, M{\"u}nster,
  Sharma, Sheinman, Jawerth, Fabry, Weitz, and MacKintosh}}]{Licup2015}
\bibinfo{author}{\bibfnamefont{A.~J.} \bibnamefont{Licup}},
  \bibinfo{author}{\bibfnamefont{S.}~\bibnamefont{M{\"u}nster}},
  \bibinfo{author}{\bibfnamefont{A.}~\bibnamefont{Sharma}},
  \bibinfo{author}{\bibfnamefont{M.}~\bibnamefont{Sheinman}},
  \bibinfo{author}{\bibfnamefont{L.~M.} \bibnamefont{Jawerth}},
  \bibinfo{author}{\bibfnamefont{B.}~\bibnamefont{Fabry}},
  \bibinfo{author}{\bibfnamefont{D.~A.} \bibnamefont{Weitz}}, \bibnamefont{and}
  \bibinfo{author}{\bibfnamefont{F.~C.} \bibnamefont{MacKintosh}},
  \bibinfo{journal}{arXiv.org} p. \bibinfo{pages}{1503.00924}
  (\bibinfo{year}{2015}).

\bibitem[{\citenamefont{Sharma et~al.}(2015)\citenamefont{Sharma, Licup, Rens,
  Sheinman, Jansen, Koenderink, and MacKintosh}}]{Sharma}
\bibinfo{author}{\bibfnamefont{A.}~\bibnamefont{Sharma}},
  \bibinfo{author}{\bibfnamefont{A.~J.} \bibnamefont{Licup}},
  \bibinfo{author}{\bibfnamefont{R.}~\bibnamefont{Rens}},
  \bibinfo{author}{\bibfnamefont{M.}~\bibnamefont{Sheinman}},
  \bibinfo{author}{\bibfnamefont{K.~A.} \bibnamefont{Jansen}},
  \bibinfo{author}{\bibfnamefont{G.~H.} \bibnamefont{Koenderink}},
  \bibnamefont{and} \bibinfo{author}{\bibfnamefont{F.~C.}
  \bibnamefont{MacKintosh}}, \bibinfo{journal}{arXiv.org} p.
  \bibinfo{pages}{1506.07792} (\bibinfo{year}{2015}), \eprint{1506.07792}.

\end{thebibliography}

\end{document}